\documentclass{webofc}
\usepackage[varg]{txfonts}   % Web of Conferences font
\usepackage{graphicx}
\usepackage{amsmath}
\newcommand{\beq}{\begin{equation}}
\newcommand{\eeq}{\end{equation}}
\newcommand{\beqs}{\begin{eqnarray}}
\newcommand{\eeqs}{\end{eqnarray}}

\newcommand\Fig[1]{Fig.~\ref{fig:#1}}
\newcommand\Eq[1]{Eq.~(\ref{eq:#1})}

\begin{document}
\title{Composite dynamics in Sp($2N$) gauge theories}

\author{\firstname{Jong-Wan} \lastname{Lee}\inst{1,2}\fnsep\thanks{\email{jwlee823@pusan.ac.kr}} \and
        \firstname{Ed} \lastname{Bennett}\inst{3} \and
        \firstname{Deog Ki} \lastname{Hong}\inst{1} \and
        \firstname{Ho} \lastname{Hsiao}\inst{4}\and
        \firstname{C.-J. David} \lastname{Lin}\inst{4,5}\and
        \firstname{Biagio} \lastname{Lucini}\inst{3,6}\and
        \firstname{Maurizio} \lastname{Piai}\inst{7}\and
        \firstname{Davide} \lastname{Vadacchino}\inst{8}
        % etc.
}
%\author{\firstname{Jong-Wan} \lastname{Lee}\inst{1,2}\fnsep\thanks{\email{jwlee823@pusan.ac.kr}} \and
%        \firstname{Ed} \lastname{Bennett}\inst{3}\fnsep\thanks{\email{e.j.bennett@swansea.ac.uk}} \and
%        \firstname{Deog Ki} \lastname{Hong}\inst{1}\fnsep\thanks{\email{dkhong@pusan.ac.kr}} \and
%        \firstname{Ho} \lastname{Hsiao}\inst{4}\fnsep\thanks{\email{thepaulxiao@gmail.com}} \and
%        \firstname{C.-J. David} \lastname{Lin}\inst{4,5}\fnsep\thanks{\email{dlin@nycu.edu.tw}} \and
%        \firstname{Biagio} \lastname{Lucini}\inst{3,6}\fnsep\thanks{\email{b.lucini@swansea.ac.uk}} \and
%        \firstname{Maurizio} \lastname{Piai}\inst{7}\fnsep\thanks{\email{m.piai@swansea.ac.uk}} \and
%        \firstname{Davide} \lastname{Vadacchino}\inst{8}\fnsep\thanks{\email{davide.vadacchino@plymouth.ac.uk}}
%        % etc.
%}
\institute{Department of Physics, Pusan National University, Busan 46241, Korea
\and
           Institute for Extreme Physics, Pusan National University, Busan 46241, Korea
\and
           Swansea Academy of Advanced Computing, Swansea University (Bay Campus), Fabian Way, SA1 8EN Swansea, Wales, United Kingdom
\and
	   Institute of Physics, National Yang Ming Chiao Tung University, 1001 Ta-Hsueh Road, Hsinchu 30010, Taiwan
\and
	   Center for High Energy Physics, Chung-Yuan Christian University, Chung-Li 32023, Taiwan
\and
	   Department of Mathematics, Faculty of Science and Engineering, Swansea University (Bay Campus), Fabian Way, SA1 8EN Swansea, Wales, United Kingdom
\and
	   Department of Physics, Faculty of Science and Engineering, Swansea University (Park Campus), Singleton Park, SA2 8PP Swansea, Wales, United Kingdom
\and
	   Centre for Mathematical Science, University of Plymouth, Plymouth, PL4 8AA, United Kingdom
          }

\abstract{%
Sp($2N$) gauge theories with fermonic matter provide an ideal laboratory to build extensions 
of the standard model based on novel composite dynamics. 
Examples include composite Higgs along with top partial compositeness and composite dark matter. 
Without fermions, their study also complements those based on SU($N_c$) gauge theories 
with which they share a common sector in the large $N_c=2N$ limit. 
We report on our recent progress in the numerical studies of Sp($2N$) gauge theories discretised 
on a four-dimensional Euclidean lattice. 
In particular, we present preliminary results for the low-lying mass spectra of mesons and chimera baryons in the theories with $N=2$. 
We also compute the topological susceptibility for various values of $N$, extrapolate the results to the large $N$ limit, 
and discuss certain universal properties in Yang-Mills theories.
}
\maketitle
%
%%%%%%%%%%%%%%%%%%%%%%%%%%%%%%%%%%%%%%%%%%%%%%%%%%%%%%%%%%%%%%%%%%%%%%%%%%%%%
\section{Introduction}
\label{sec:intro}

%basic properties of Sp(2N) gauge theories\\
%theoretical interests - supplement to SU(N)\\
%phenomenological interests - composite Higgs, top partial compositeness, dark matter, gravitational waves\\
%particular model with Sp(4) + theory space

Non-abelian gauge theories based on Sp($N_c=2N$) groups coupled to fermions exhibit interesting nonperturbative phenomena, 
such as confinement and global symmetry breaking, analogous to the SU($N_c$) gauge theories. 
In Yang-Mills theories, they provide another family of groups in which certain universal features are expected to emerge in the large $N_c$ limit. 
In the opposite extreme limit we have SU($2$) $=$ Sp($2$). A distinctive feature is that the fundamental representation of Sp($2N$) is pseudoreal. 
When the gauge dynamics is coupled to $N_f$ fundamental Dirac flavours, the main consequences are two-fold. 
On the one hand, finite density calculations are free from the {\it sign problem}, a notorious hindrance in the numerical studies of QCD, 
and thus may provide new physical insights on the phase diagram. 
On the other hand, the global symmetry associated with the flavour is enhanced to SU($2N_f$), 
which is spontaneously and explicitly broken into Sp($2N_f$) in the presence of fermion condensation and non-zero (degenerate) mass. 
Because such a symmetry breaking pattern yields a comparatively large coset space, 
one can take advantage of it when constructing certain classes of phenomenological extensions of the standard model (SM), 
%based on novel strongly coupled gauge theories, 
such as composite Higgs, top partial compositeness, and a strongly interacting massive particle (SIMP) as dark matter candidate. 
Furthermore, it is known that the finite-temperature transition of pure Sp($2N$) gauge theories is first order for $N\geq 2$ \cite{Holland:2003kg}, 
which has a potential impact on studies of the cosmological evolution of the early universe and on gravitational wave experiments. 
It should be emphasized that, in principle, Sp($2N$) gauge theories are equally important as SU($N_c$) in the searches for new physics 
based on novel strongly coupled gauge theories.

\begin{figure}[t]
%For one-column wide figures use syntax of figure~\ref{fig-1}
\centering
\includegraphics[width=0.5\textwidth]{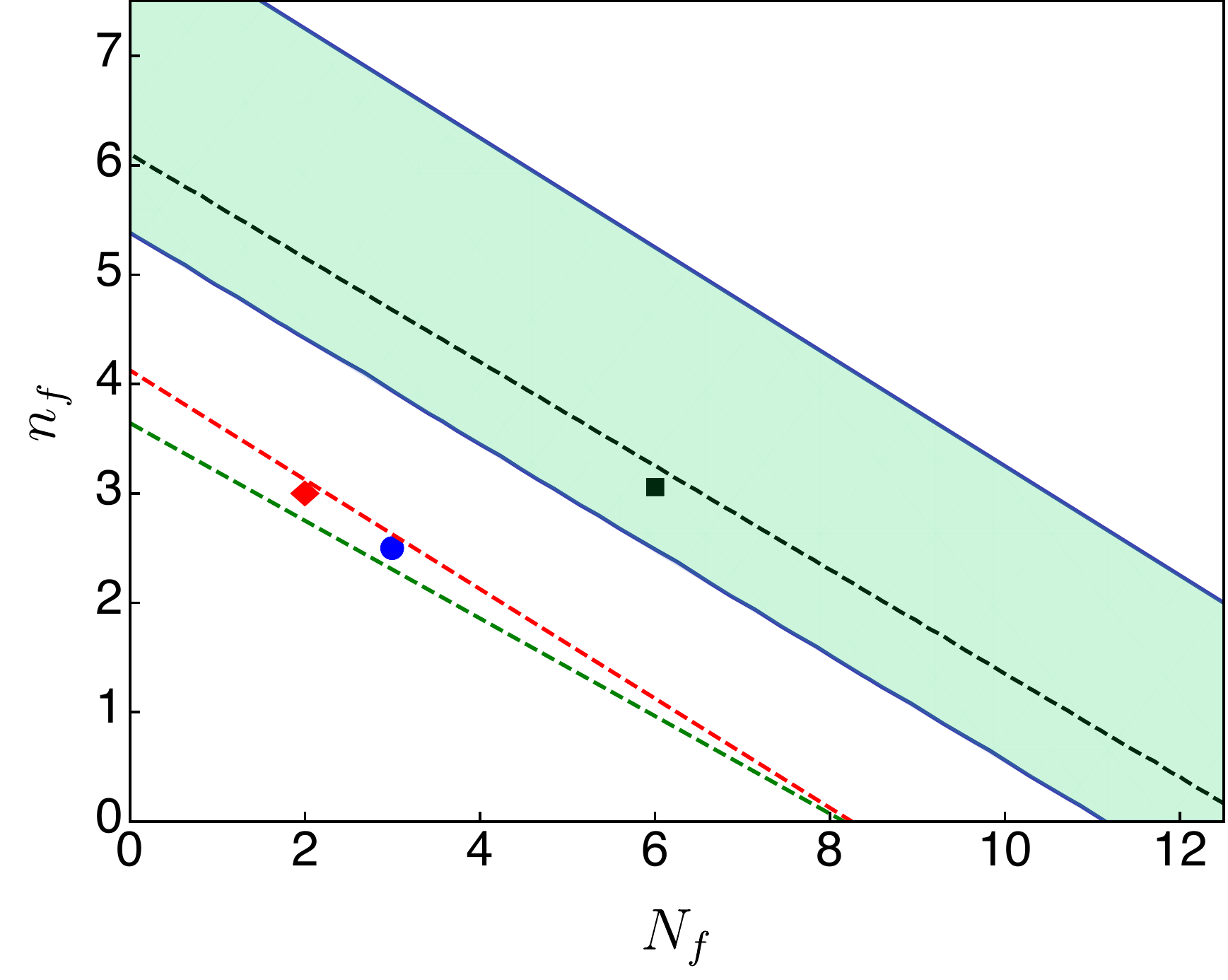}
\caption{Phase diagram of Sp($4$) gauge theory with $N_f$ fundamental and $n_f$ antisymmetric Dirac fermions. 
The shaded region denotes the conformal window recently estimated in Ref.~\cite{Kim:2020yvr}, 
while dashed lines denote other analytical estimations in the literature, which are discussed in the same reference. 
The coloured dots represent the theories realising the phenomenological models discussed in the main text. 
}
\label{fig:phasesp4}       % Give a unique label
\end{figure}

To perform lattice simulation, we need to choose a concrete model, well motivated by both theoretical and phenomenological arguments. 
We concentrate on the minimal Sp($4$) theory with which 
one can build both composite Higgs and top partial compositeness. 
In \Fig{phasesp4}, we show the theory space of Sp($4$) with $N_f$ fundamental and $n_f$ $2$-index antisymmetric Dirac fermions. 
Representative models taken from the literature are denoted by red \cite{Barnard:2013zea,Ferretti:2016upr}, blue \cite{Ferretti:2016upr} and black \cite{Cacciapaglia:2019dsq} colours. 
As the infrared (IR) behaviour of strongly coupled gauge theories determines what type of applications they suit, 
in the figure we also present the results of a recent analytical study of the edge of the conformal window (CW), 
compared with other analytical estimations 
(see Ref.~\cite{Kim:2020yvr} and references therein for the details). 

The theory with $N_f=2$ fundamental and $n_f=3$ antisymmetric Dirac fermions 
is expected to be in the chirally broken phase. The resulting $5$ pseudo Nambu-Goldstone bosons (pNGBs) in the fundamental sector 
include the SM Higgs doublets. This theory can be studied by employing currently available numerical lattice techniques without altering the flavour structure 
of the original, phenomenologically motivated model. 
Since numerical studies of gauge theories with mixed representations are relatively new to the lattice community, 
and computationally demanding in the presence of dynamical fermions, 
we have first considered the quenched approximation \cite{Bennett:2017kga,Bennett:2019cxd} 
and in stages introduced dynamical Dirac flavours in the form of two fundamental \cite{Bennett:2019jzz}, 
three antisymmetric \cite{Lee:2022elf} and both combined \cite{Bennett:2022yfa}. 

In parallel, we have been studying pure Sp($2N$) gauge theories on the lattice with various values of $N$, 
$N=1,\,2,\,3,\,4$, and furthermore considered the extrapolation towards the large $N$ limit. 
We have studied the string tension, the mass spectra of glueballs with many different quantum numbers (including some excited states) \cite{Bennett:2020qtj}, 
the Wilson flow and the topological susceptibility \cite{Bennett:2022ftz}, 
%, and certain ratios conjectured to be universal
and found empirical evidence of the emergence of universal patterns in Yang-Mills theories \cite{Hong:2017suj,Bennett:2020hqd,Bennett:2022gdz}. 

%%%%%%%%%%%%%%%%%%%%%%%%%%%%%%%%%%%%%%%%%%%%%%%%%%%%%%%%%%%%%%%%%%%%%%%%%%%%%
\section{Lattice setup}
\label{sec:lattice}

The lattice action of Sp($2N$) gauge theories is discretised in a four-dimensional Euclidean space with temporal and spatial extents  $T$ and $L$, respectively. 
We adopt the standard plaquette action for pure gauge interactions 
and the Wilson-Dirac action for gauge-fermion interactions. Details can be found in Refs.~\cite{Bennett:2022ftz,Bennett:2022yfa} and references therein. 
Numerical calculations have been carried out by employing the Heat Bath algorithms for the pure Sp($2N$) theories 
and the (rational) hybrid Monte Carlo algorithms for the Sp($4$) theories with dynamical fermions. 
The lattice action contains a lattice coupling $\beta \equiv 4 N/g^2$ and fermion masses $am_0^f$ and $am_0^{as}$ 
for the fundamental and antisymmetric representations, respectively. 
Here, $g$ is the bare gauge coupling and $a$ is the lattice spacing. 

The lattice parameters need to be chosen carefully so that the lattice measurements can be extrapolated 
to the corresponding continuum ones by taking $\beta\rightarrow\infty$ in a controlled way. 
In particular, we invegstigated the presence of a first-order bulk phase transition 
and, if it exists, determined the critical coupling $\beta^c$ associated with the boundary between strong and weak coupling regimes. 
We have shown that Sp($2N$) Yang-Mills for $N=1,\,2,\,3,\,4$ exhibits a smooth crossover \cite{Bennett:2020qtj}, 
while dynamical Sp($4$) theories undergo a first-order transition in the strong coupling regime \cite{Bennett:2017kga,Lee:2018ztv,Bennett:2022yfa}. 
Having the values of $\beta^c$ in hand, we have performed all the numerical simulations in the weak coupling regime, $\beta> \beta^c$, 
and extrapolated to the continuum, if possible. 
Furthermore, we have used choices of the lattice volume for which finite size corrections to the observables are statistically negligible. 
%To analyse numerical results for the lattice theories with different bare parameters in the same physical reference, 
We elected to set a scale for dimensionful quantities by using the gradient flow method \cite{Luscher:2010iy}. 
Two different definitions of the flow scale, $t_0$ \cite{Luscher:2010iy} and $w_0$ \cite{Borsanyi:2012zs}, are available. 
Furthermore, the string tension $\sigma$ can be used in the case of pure Sp($2N$) Yang-Mills.

%%%%%%%%%%%%%%%%%%%%%%%%%%%%%%%%%%%%%%%%%%%%%%%%%%%%%%%%%%%%%%%%%%%%%%%%%%%%%
\section{Results: Sp($2N$) Yang-Mills}
\label{sec:pursp}

In this section, we briefly summarise our main findings on Sp($N_c=2N$) pure gauge theories with $N=1,\,2,\,3,\,4$ and discuss certain universal features in Yang-Mills theories 
by comparing to SU($N_c$) and SO($N_c$) gauge theories. 
We refer the reader to Refs.~\cite{Bennett:2020qtj,Bennett:2022ftz,Bennett:2020hqd,Bennett:2022gdz} for the full details of lattice technicalities and numerical results. 

Two main observables are the mass of glueballs and the string tension, which are extracted from the large-time behaviour of the $2$-point correlation functions 
built out of Wilson and Polyakov loops, respectively. 
%The latter basically gives rise to a linear potential between static quark and antiquark, a characteristic feature of confinement, 
%and the string tension has been determined by examining the $L$ dependence of the numerical data. 
Two-point functions involving Polyakov loops have been measured to extract the string tension. 
The masses of glueballs are obtained by studying a variational problem involving correlators between two Wilson loops of various shapes and sizes. 
The spin quantum numbers $J$ of the glueball states in the continuum theory 
are determined through the decomposition of the irreducible representations of the octahedral group---the symmetry group of a cubic lattice---in the lattice theory. 
%while the parity $P$ and the charge conjugation $C=+1$ are identified in the same way. 
Using variational techniques, we were able to extract the masses for the ground states with $J^P=0^{\pm},\,1^{\pm},\,2^{\pm},\,3^{\pm}$ 
as well as for the first excited states with $0^{\pm}$ \cite{Bennett:2020qtj}. 
After performing the continuum extrapolations of the glueball masses in units of $\sqrt{\sigma}$ at each $N_c$, 
we took the large $N_c$ limit by including a $\mathcal{O}(1/N_c)$ correction term. 
We found that the resulting values at $N_c=\infty$ are in good agreement with those for SU($N_c$) \cite{Bennett:2020qtj,Lucini:2021xke}. 
%supporting the large-$N$ universality in Yang-Mills theories. 
We also note that the lightest and the second lightest states are the positive-parity scalar and tensor glueballs. 
%Combining the results for $N=2,\,3,\,4$ in units of $\sqrt{\sigma}$, we 

%glueballs, string tension, and susceptibility\\
%universal features in Yang-Mills 

%paragraph a unique label (see Sect.~\ref{sec-1}).

%For two-column wide figures use syntax of figure~\ref{fig-2}
\begin{figure}[t]
\centering
%\vspace*{5cm}       % Give the correct figure height in cm
\includegraphics[width=0.45\textwidth]{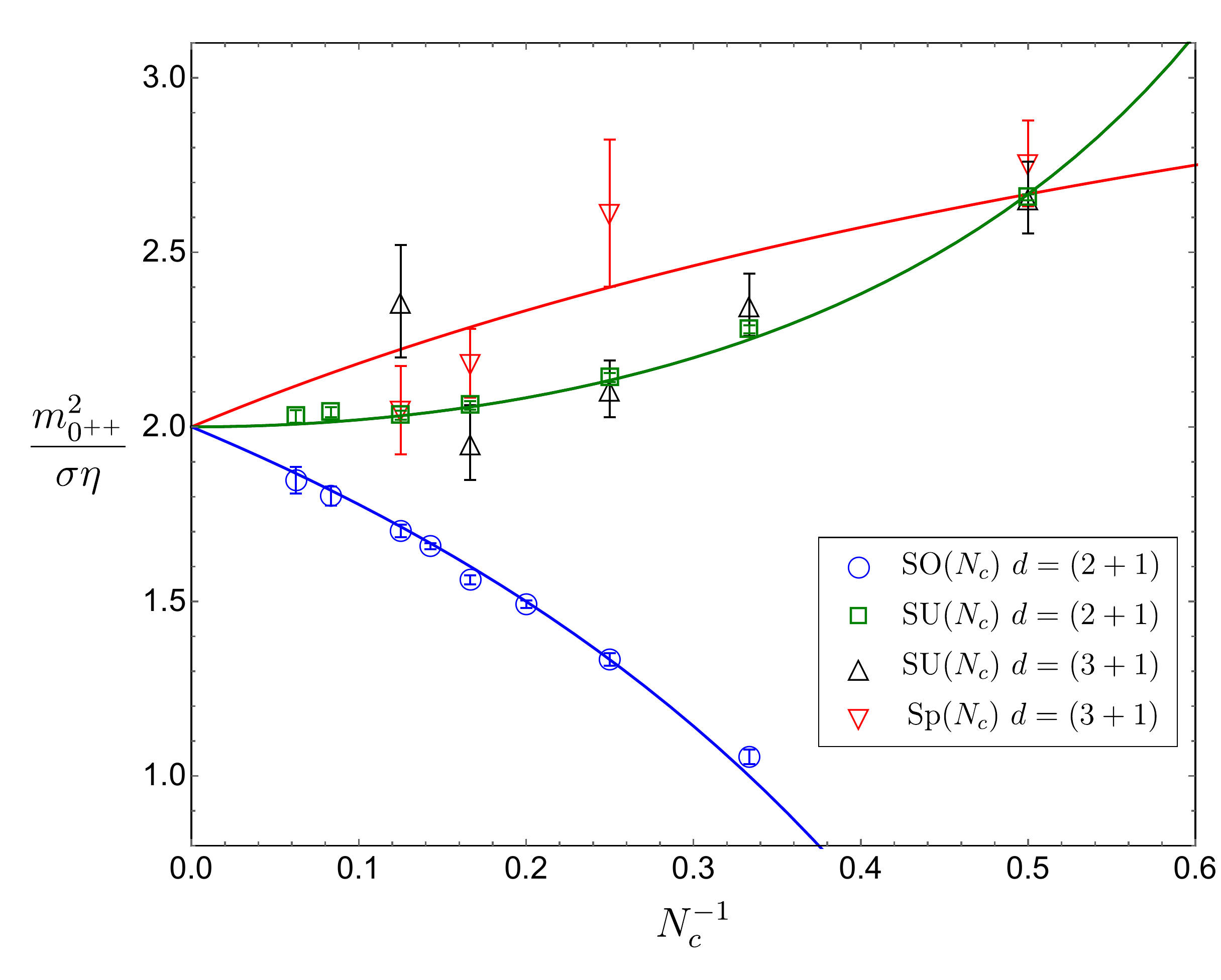}
\hspace{0.8cm}
\includegraphics[width=0.4\textwidth]{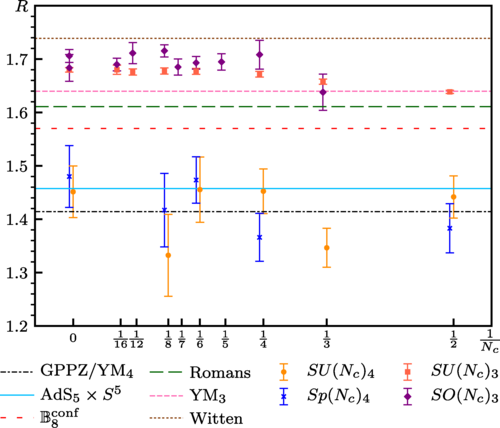}
\caption{Numerical results and extrapolations towards the large $N_c$ limit, 
showing the convergence of the SU($N_c$), Sp($N_c$) and SO($N_c$) Yang-Mills theories: 
(left) Casimir scaling in the lightest scalar glueballs, 
%\cite{Hong:2017suj,Bennett:2020qtj}, 
and (right) mass ratio between the lightest scalar and tensor glueballs, $R\equiv m_{2^{++}}/m_{0^{++}}$. 
% taken from Ref.~\cite{Bennett:2020hqd}. 
The constant $\eta$ is defined in \Eq{casimir}. 
Figures from Refs.~\cite{Hong:2017suj,Bennett:2020qtj,Bennett:2020hqd}. 
}
\label{fig:casirmir}       % Give a unique label
\end{figure}

Recently, it has been argued that certain ratios between physical observables, up to appropriate gauge-group factors, 
might exhibit universal behaviours even at finite $N_c$ in SU($N_c$), Sp($N_c$) and SO($N_c$) Yang-Mills \cite{Hong:2017suj,Athenodorou:2016ndx}. 
Here, {\it universal} means that the ratio only depends on the space-time dimension of the theory, $d$, regardless of the number of colours $N_c$ and the gauge groups. 
First of all, in Ref.~\cite{Hong:2017suj} the authors conjectured that the ratio $m_{0^{++}}^2/\sigma$ is proportional to 
the ratio of eigenvalues of quadratic Casimir operators in the adjoint (Adj) and the fundamental (F) representations, 
which is supported by existing lattice results for SO($N_c$) in $d=(2+1)$ and SU($N_c$) in both $d=(2+1)$ and $(3+1)$ theories. 
Combining our results for Sp($N_c$), as shown in \Fig{casirmir}, 
we obtained the following results for the universal constant $\eta$ \cite{Hong:2017suj,Bennett:2020qtj}
\beq
\eta \equiv \frac{m_{0^{++}}^2}{\sigma}\cdot\frac{C_2(F)}{C_2(Adj)} = 
\begin{cases}
    5.388(81)(60), & d=(3+1),\\
    8.440(14)(76), & d=(2+1),
  \end{cases}
\label{eq:casimir}
\eeq
where the first and second parentheses denote the statistical and systematic errors, respectively. 

A second interesting quantity is the mass ratio between the lightest parity-even scalar and tensor glueballs, 
$R\equiv m_{2^{++}}/m_{0^{++}}$, first proposed in Ref.~\cite{Athenodorou:2016ndx}.  
In \Fig{casirmir}, we present the lattice results available in the literature as well as analytical results obtained by gauge-gravity dualities and field-theoretical calculations. 
Details are found in Ref.~\cite{Bennett:2020hqd}. 
As seen in the figure, it is evident that the ratio $R$ is independent of $N_c$ and the gauge group, but gives rise to different values for $d=(2+1)$ and $(3+1)$ dimensions. 

\begin{figure}[h]
\centering
%\sidecaption
\includegraphics[width=0.45\textwidth]{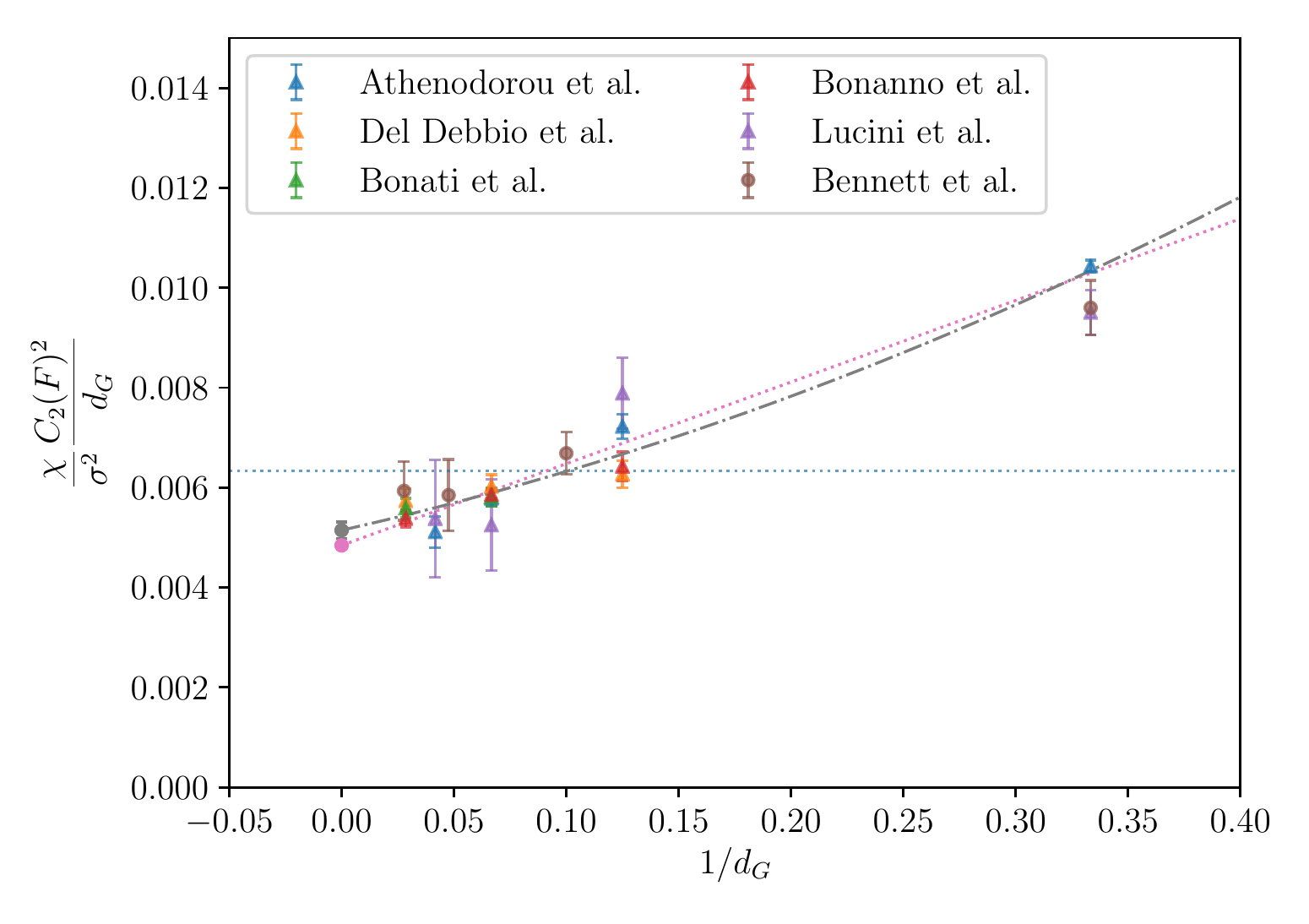}
\caption{Topological susceptibilities $\chi$ in units of the string tension $\sigma$, rescaled by group factors $C_2(F)^2/d_G$, 
for SU($N_c$) and Sp($N_c$) Yang-Mills as a function of the dimension of the group $d_G$. 
The figure is taken from Ref.~\cite{Bennett:2022gdz}. 
}%
\label{fig:suscept}       % Give a unique label
\end{figure}

%Beyond the spectral results of Yang-Mills, 
An interesting non-perturbative quantity, which is closely related with the $U(1)_A$ problem and the $\theta$-term, 
is the topological susceptibility $\chi$. Since $\chi$ is defined as a second derivative of the vacuum density $F(\theta)$ with respect to $\theta$, and 
each gauge field contributes equally, it has been conjectured to be proportional to the dimension of group $d_G$. 
We hence investigate the behaviour of the ratio $\eta_\chi \equiv \chi C_2(F)^2/\sigma^2 d_G$ \cite{Bennett:2022gdz}. 
A survey of lattice results for $d=(3+1)$ SU($N_c$) and Sp($N_c$) Yang-Mills can be found in \Fig{suscept}.  
While the two sequences of gauge groups agree with each other, we observe a visible dependence on $d_G$. 
After accounting for the $N_c$ dependence, we took the large-$N_c$ extrapolation and obtained \cite{Bennett:2022gdz}: 
%$\eta_\chi(\infty) = (48.42\pm 0.77\pm3.31)\times 10^{-4}$ \cite{Bennett:2022gdz}.
\beq
\eta_\chi(N_c=\infty) = 48.4(8)(33)\times 10^{-4},
\eeq
where the first and second parentheses again denote the statistical and the systematic errors. 

%For tables use syntax in table~\ref{tab-1}.
%\begin{table}
%\centering
%\label{tab:tab1}       % Give a unique label
%% For LaTeX tables you can use
%\begin{tabular}{lll}
%\hline
%first & second & third  \\\hline
%number & number & number \\\hline
%\end{tabular}
%% Or use
%\vspace*{5cm}  % with the correct table height
%\end{table}
%%

%%%%%%%%%%%%%%%%%%%%%%%%%%%%%%%%%%%%%%%%%%%%%%%%%%%%%%%%%%%%%%%%%%%%%%%%%%%%%
\section{Results: Sp($4$) with fermions}
\label{sec:pursp}

\begin{figure}[t]
\centering
\sidecaption
\includegraphics[width=0.41\textwidth]{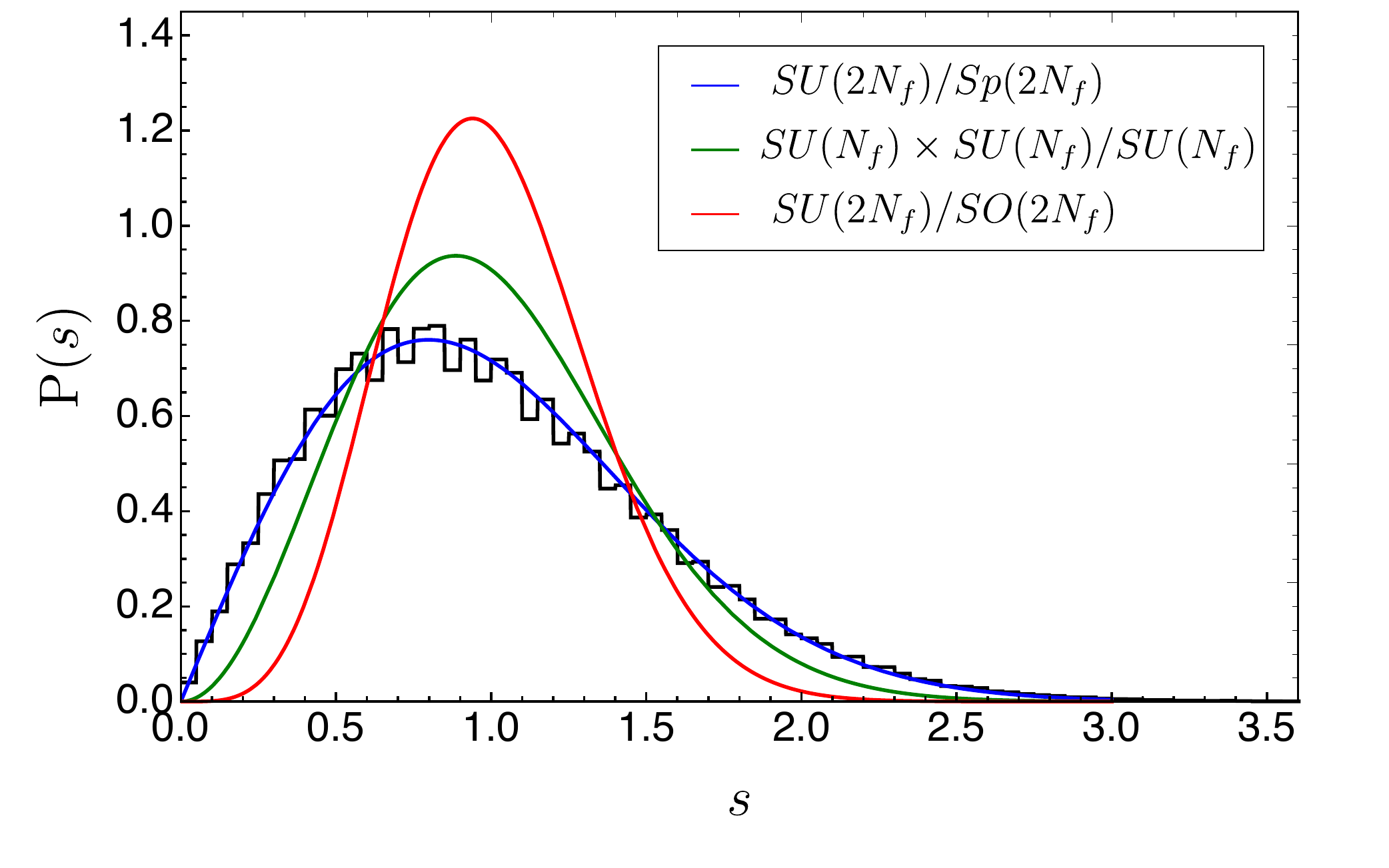}
\hspace{0.7cm}
\includegraphics[width=0.41\textwidth]{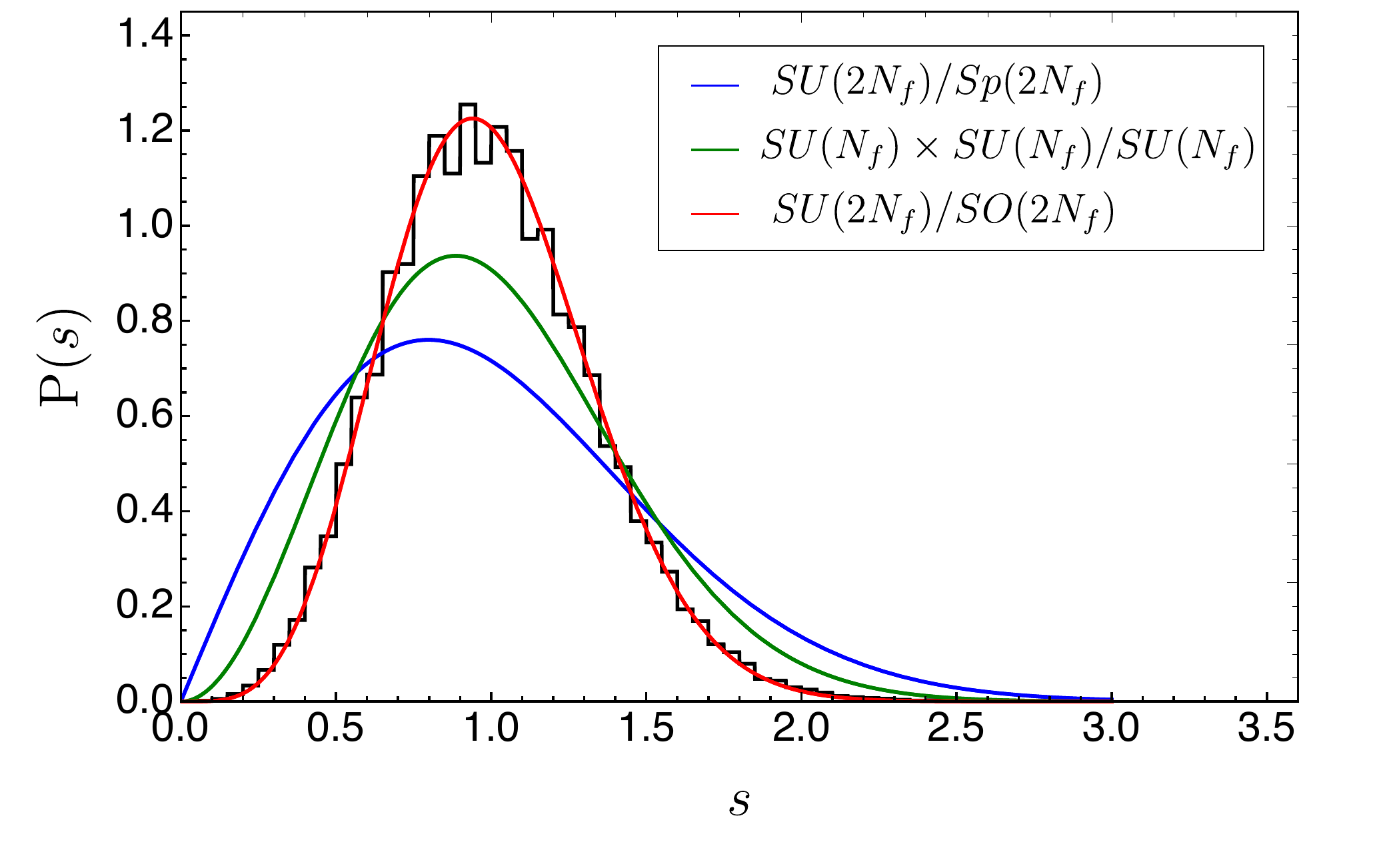}
\caption{The unfolded density of the spacings between subsequent Dirac eigenvalues for the fundamental (left) and antisymmetric (right) representations. 
The black histogram denotes the numerical data, while the coloured solid lines are the $\chi$RMT predictions. 
Figures from Ref.~\cite{Bennett:2022yfa}.
}
\label{fig:symbreaking}       % Give a unique label
\end{figure}

%As mentioned in \Sec{intro}, 
We restrict ourselves to the Sp($4$) gauge group for the cases involving fermionic matter fields. 
Because these are the first numerical studies of fermions in the fundamental and antisymmetric representations transforming under Sp($2N$) with $N\geq 2$, 
we have performed nontrivial tests of the algorithms. 
A particularly important cross-check pertains the breaking pattern of the flavour symmetry. 
%To do this, 
Following the work in Ref.~\cite{Cossu:2019hse}, we calculated the eigenvalues of the Dirac operators in the quenched ensemble on a small lattice of size $4^4$, 
and compared the unfolded density of spacings, $P(s)$, with the predictions of chiral random matrix theory ($\chi$RMT) \cite{Bennett:2022yfa}. 
As seen in \Fig{symbreaking}, the histograms of our numerical data are well described by the $\chi$RMT predictions, denoted as solid lines, 
with the expected breaking patterns of SU($2N_f$)/Sp($2N_f$) and SU($2N_f$)/SO($2N_f$) for the fundamental and antisymmetric fermions, respectively. 
%\cite{Cossu:2019hse}

%symmetry breaking patterns - Dirac spectrum, mass and decay constant of pseudoscalar mesons

We now turn our attention to the results for dynamical fermions. 
In Ref.~\cite{Bennett:2019jzz}, we reported the spectral results for flavoured mesons in the lightest spin-$0$ and -$1$ channels, 
measured on dynamical ensembles with $N_f=2$ fundamental Dirac fermions. 
We examined the fermion mass dependence of the mass and decay constant of pseudoscalar meson, the lightest state in the spectrum, 
and found clear evidence of the spontaneous breaking of the global symmetry: 
the mass squared is proportional to the fermion mass and the decay constant extrapolates to a non-zero value in the massless limit. 
This is expected by the fact that the pNGBs span the
%Such a result strongly supports the model buildings based on 
SU($4$)/Sp($4$) coset relevant for models of composite Higgs and SIMP dark matter.
Concerning the theory with $n_f=3$ antisymmetric Dirac fermions, we also found some evidence of 
the spontaneous beaking of the global symmetry in our preliminary results \cite{Lee:2022elf}, 
which need to be further confirmed by extending the calculations to smaller fermion mass. 
In the case of a fully dynamical Sp($4$) theory with both two fundamental and three antisymmetric fermions, 
we have carried out lattice studies on a single ensemble so far \cite{Bennett:2022yfa}, 
and thus cannot yet make any statement on its IR nature. 

%vector meson in neighboring gauge theories with two Dirac flavours

\begin{figure}[t]
\centering
\sidecaption
\includegraphics[width=0.41\textwidth]{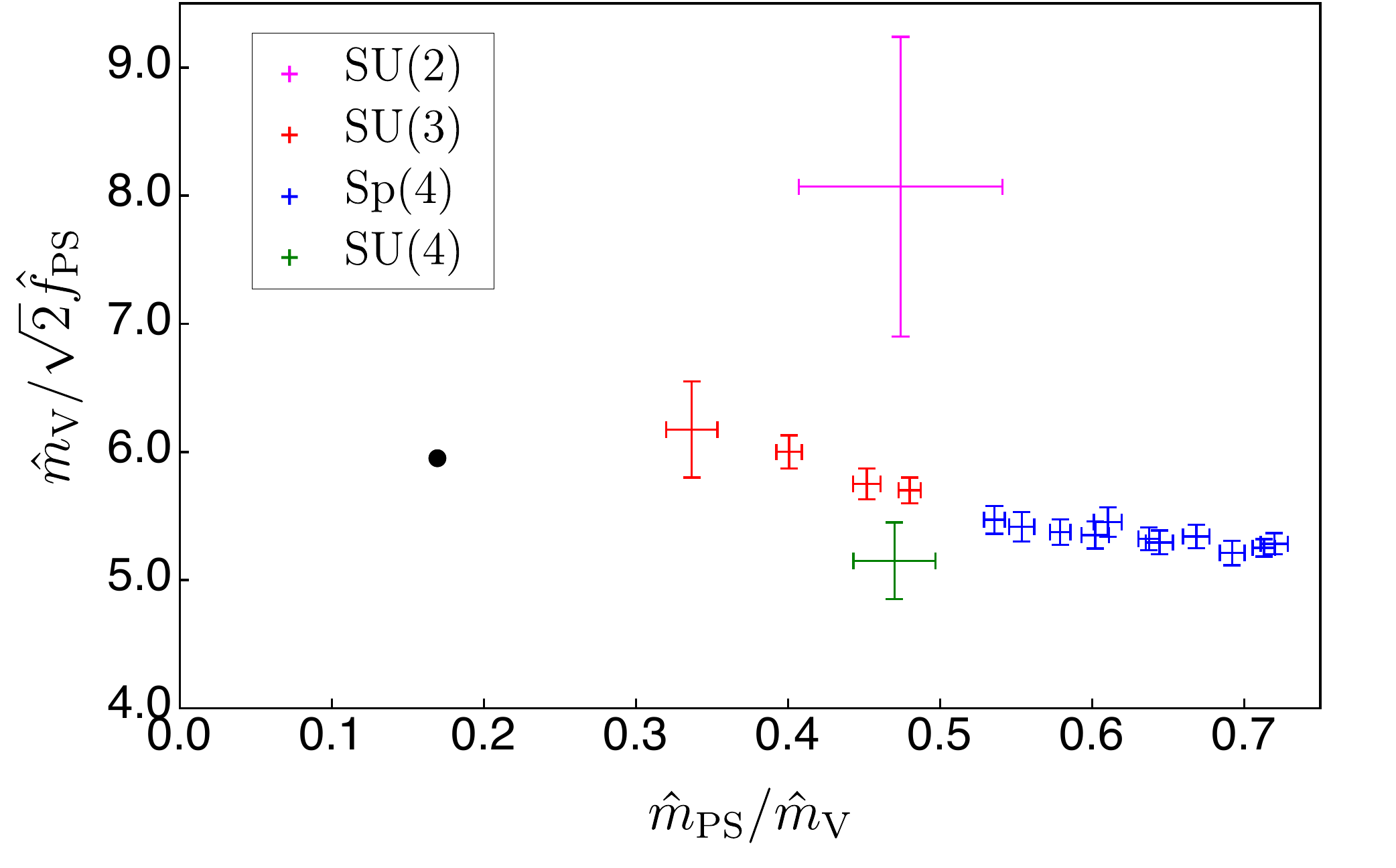}
\hspace{0.7cm}
\includegraphics[width=0.41\textwidth]{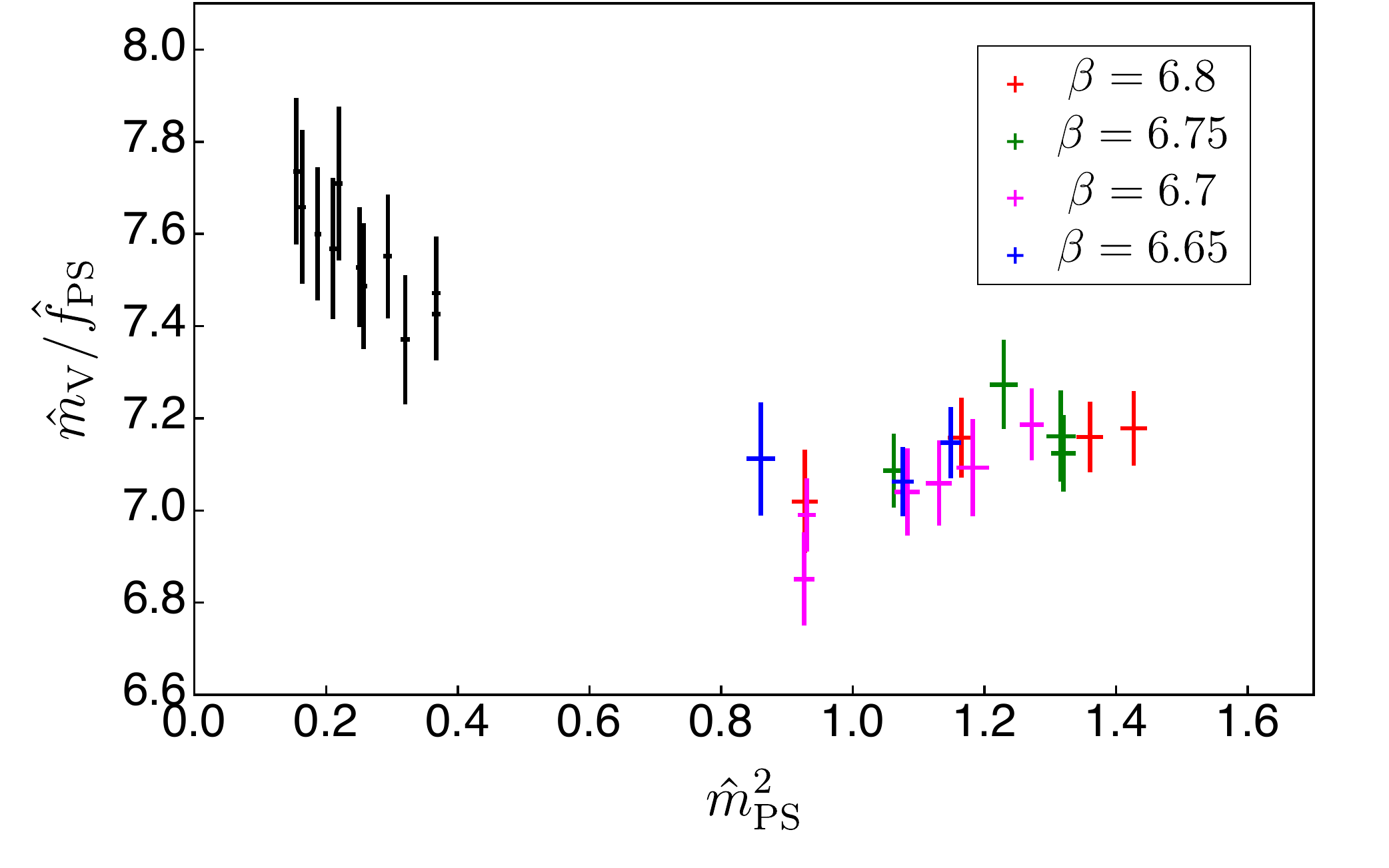}
\caption{(left) Vector meson masses in various gauge theories with $N_f=2$ fundamental Dirac flavours: 
%taken from Ref.~\cite{Bennett:2019jzz}: 
different colours are for various gauge groups, while the black dot denotes the QCD value. 
%magenta, red, blue, and green colours denote 
%SU(2), SU(3), Sp(4) and SU(4) theories, while the black dot denotes the QCD value \cite{Bennett:2019jzz}. 
(right) Vector meson masses in Sp($4$) theory with $N_f=2$ fundamental (in the continuum, black) and $n_f=3$ antisymmetric 
(colours for different $\beta$ values) Dirac flavours. 
The hatted notation denotes the mass in units of the gradient flow scale, $w_0$, e.g. $\hat{m}_{\rm PS} = m_{\rm PS} w_0$. 
Figures are taken from Refs.~\cite{Bennett:2019jzz,Lee:2022elf}. 
}
% taken from Ref.~\cite{Lee:2022elf}.}
\label{fig:vectormass}       % Give a unique label
\end{figure}

The lightest state in the spin-$1$ channel is the vector meson. 
Its mass is expected to be non-zero in the massless limit. 
% and can be considered as the typical scale of the global symmetry breaking if exists. 
A particularly interesting quantity is its mass in units of the pseudoscalar decay constant, $m_{\rm V}/\sqrt{2}f_{\rm PS}$, 
which is closely related with the low-energy constant associated with its decay to two pseudoscalar mesons via the phenomenological KSRF relation 
\cite{Kawarabayashi:1966kd,Riazuddin:1966sw}. 
In \Fig{vectormass}, we show our results for the $N_f=2$ fundamental Sp($4$) theory in the continuum limit, compared to other gauge theories \cite{Bennett:2019jzz}. 
The resulting masses are below the threshold of the two-pseudoscalar decay, 
yet our results indicate a value for the ratio in 
%but it may not be difficult to find that 
Sp($4$) that is somewhere between SU($3$) and SU($4$) theories near the threshold. 

The right panel of \Fig{vectormass} shows our preliminary results for the same ratio 
for the $n_f=3$ antisymmetric Sp($4$) theory measured at different $\beta$ values. 
%values of the lattice coupling.  
For comparison, we also present the continuum results of the $N_f=2$ fundamental Sp($4$) theory. 
Although a direct comparison may not be possible due to the different mass range 
(and the absence so far of a continuum extrapolation in the former case), 
%in the case of antisymmetric fermions), 
we can see that the former is slightly smaller than the latter. 
This result is consistent with large-$N_c$ argument: $m_{\rm V}$ is independent on the representation while $f_{\rm PS}^2$ is proportional to the dimension of the representation. 
A survey of the lattice results of $m_{\rm V}/f_{\rm PS}$ 
for various gauge theories, e.g. different numbers of colours and flavours, and different gauge groups and fermion representations, can be found in Ref.~\cite{Nogradi:2019auv}. 
%The ratio $m_{\rm V}/\sqrt{2}f_{\rm PS}$ also provides a good testbed for the large-$N$

%chimera baryons -  mass hierarchy

\begin{figure}[t]
\centering
\sidecaption
\includegraphics[width=0.43\textwidth]{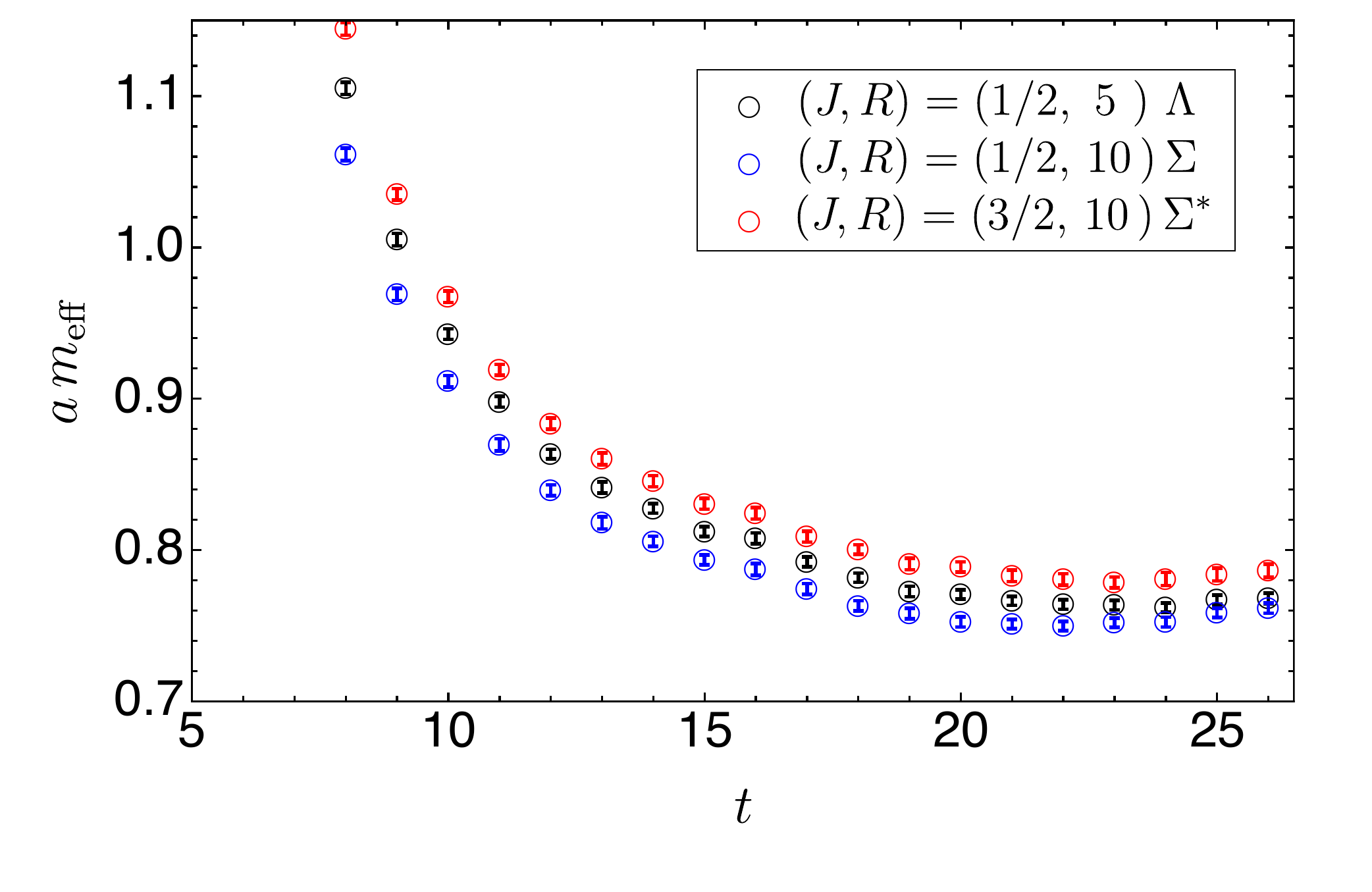}
\hspace{0.5cm}
\includegraphics[width=0.43\textwidth]{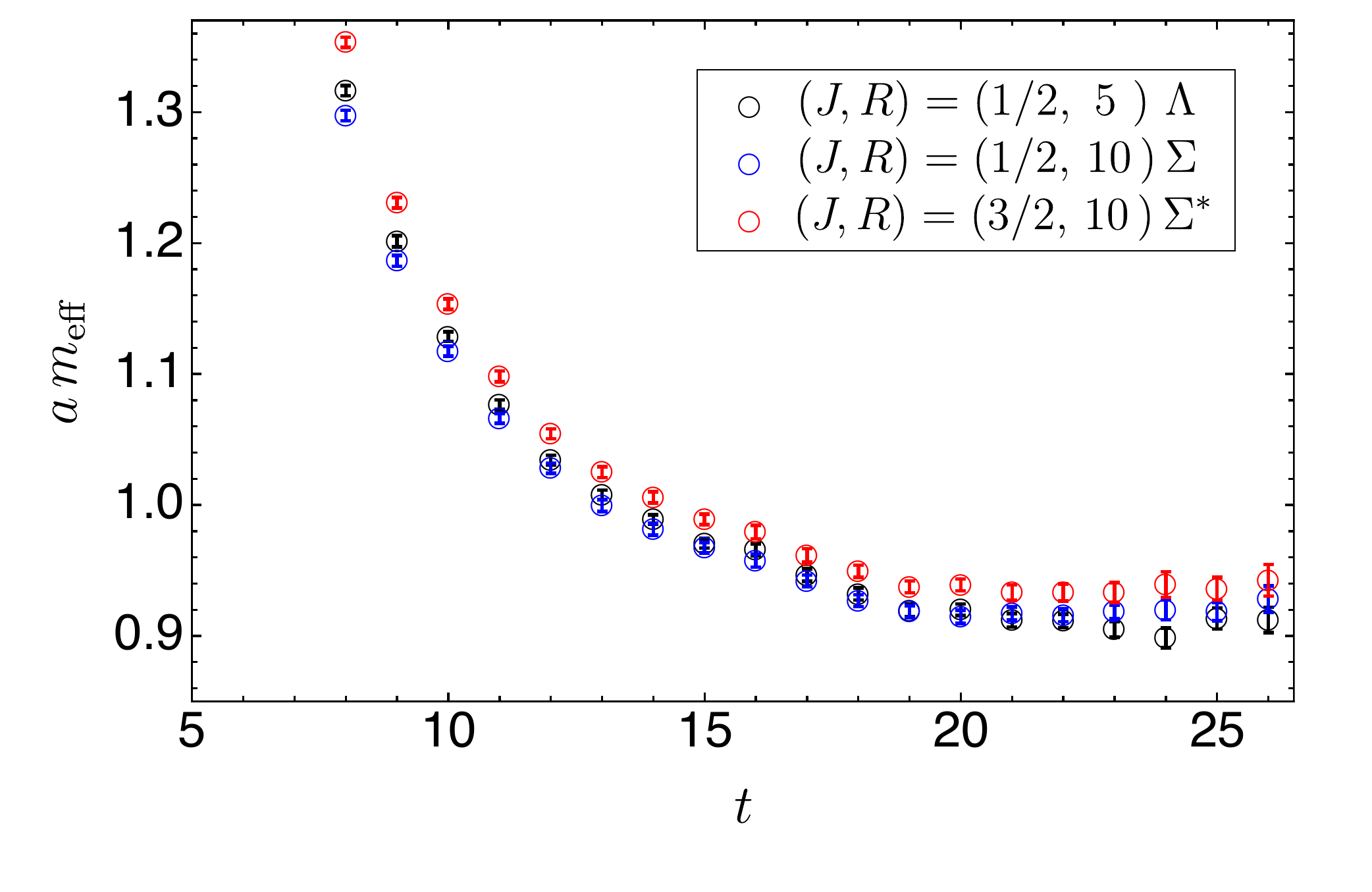}
\caption{Effective mass plots for the lightest spin-$1/2$ and $3/2$ chimera baryons in partially quenched approximation. 
%The bare parameters are $\beta=6.5$, $m_f=-0.71$, $m_{as}=-1.01$ and $54\times24^3$ \cite{Bennett:2022yfa}, 
The details of the ensemble can be found in Ref.~\cite{Bennett:2022yfa}. 
The valence fermions masses are $(am^f_v,\,am^{as}_v)=(-0.71,\,-1.01)$ and $(-0.75,\,-0.8)$ for the left and right panels, respectively. 
}
\label{fig:cb}       % Give a unique label
\end{figure}

The most interesting and distinct feature of gauge theories with mixed representations is the existence of colour-singlet baryonic objects, dubbed {\it chimera} baryons, 
some of which have the same SM quantum numbers of the top quark. 
%can play the role of top partner. 
In modern scenarios of composite Higgs, in particular, this feature is exploited to explain the large mass of the top quark, through partial compositeness. 
In the Sp($4$) theory considered in this work, the interpolating operators for the chimera baryons can be constructed 
from two fundamental and one antisymmetric fermion constituents. 
Even though the colour contraction and the flavour symmetry are completely different from QCD, 
the structure of these operators is similar to that for the QCD baryons involving one heavy quark. 
More explicitly, we label the lightest spin-$1/2$ and $3/2$ chimera states, $(J, R) = (1/2,5)$, $(1/2,10)$ and $(3/2,10)$ by $\Lambda$, $\Sigma$ and $\Sigma^*$, respectively,  
where, $J$ and $R$ denote the spin and the irreducible representation of the flavour group acting on the fundamental fermions. 

We refer the reader to Refs.~\cite{Bennett:2022yfa,Hsiao:2022kxf} for details on the interpolating operators, 
the parity and spin projections, the measurements, and exploratory studies of the spectrum. 
One observation we immediately made from our preliminary studies, using the dynamical ensemble with $(\beta,am^f_0,am^{as}_0)=(6.5,-0.71,-1.01)$, 
%$am_f=-0.71$ and $am_{as}=-1.01$ 
was that $\Lambda$, being a natural choice of top partner, e.g. Ref.~\cite{Gripaios:2009pe} (but, see also Ref.~\cite{Banerjee:2022izw} for the other possibilities), 
is not the lightest state, as shown in the left panel of \Fig{cb}. 
To further investigate the mass hierarchy of chimera baryons, we performed additional measurements in a partially quenched setup using the same ensemble---the 
valence fermion mass is different to the sea fermion, $m^{f,as}_v \neq m^{f,as}_0$. 
With the choices of $(am^f_v,am^{as}_v)=(-0.75,-0.8)$, 
%and $m_{as}^v=-0.8$ 
we find that the mass ratio between pseudoscalar mesons composed of antisymmetric and fundamental constituents $m_{\rm PS}^{as}/m_{\rm PS}^{f}$ 
is about $4.3$, and $\Lambda$ is almost degenerate with $\Sigma$ and thus the lightest state in the baryon spectrum. 
A similar behaviour has been observed in the quenched approximation \cite{Hsiao:2022kxf}. 
%We also refer the reader to Ref.~\cite{Hsiao:2022kxf} for the case in the quenched limit. 
%We note that from the point of view of phenomenological model buildings the minimal requirement for the top partial compositeness would be that the top partner is stable against the interaction of new strong sector, 
%e.g. the mass difference between $\Lambda$ and $\Sigma$ is less than the mass of the lightest pseudoscalar meson. 

%%%%%%%%%%%%%%%%%%%%%%%%%%%%%%%%%%%%%%%%%%%%%%%%%%%%%%%%%%%%%%%%%%%%%%%%%%%%%
%\section{Summary and outlook}
%\label{sec:summary}
%D
~\\
%%%%%%%%%%%%%%%%%%%%%%%%%%%%%%%%%%%%%%%%%%%%%%%%%%%%%%%%%%%%%%%%%%%%%%%%%%%%%
\begin{acknowledgement}
\noindent
{\bf Acknowledgements -} 
We thank Gabriele Ferretti for useful communications. 
The work of J.~W.~L is supported by the National Research Foundation of Korea (NRF) grant funded by the Korea government (MSIT)
(NRF-2018R1C1B3001379).
The work of E. B. has been funded in part by the UKRI Science and
Technology Facilities Council (STFC) Research Software Engineering Fellowship EP/V052489/1. 
The work of D.~K.~H. was supported by
Basic Science Research Program through the National
Research Foundation of Korea (NRF) funded by the
Ministry of Education (NRF-2017R1D1A1B06033701).
The work of H.~H. and C.~J.~D.~L. is supported
by the Taiwanese MoST Grant No. 109-2112-M-009 -006 -MY3. The work of B.~L. and M.~P. has been supported in part
by the STFC Consolidated Grants No. ST/P00055X/1 and No. ST/T000813/1. B.~L. and M.~P. received funding from
the European Research Council (ERC) under the European
Union’s Horizon 2020 research and innovation program
under Grant Agreement No. 813942. The work of B.~L. is
further supported in part by the Royal Society Wolfson
Research Merit Award No. WM170010 and by the
Leverhulme Trust Research Fellowship No. RF-2020-4619. 
The work of D.~V. is supported in part the Simons Foundation under the program “Targeted Grants to Institutes” awarded to the Hamilton Mathematics Institute.
Numerical simulations have been performed on the Swansea SUNBIRD
cluster (part of the Supercomputing Wales project) and AccelerateAI A100 GPU system,
on the local HPC clusters in Pusan National
University (PNU) and in National Yang Ming Chiao Tung University
(NYCU), and on the DiRAC Data Intensive service at Leicester. 
The Swansea SUNBIRD system and AccelerateAI are part funded
by the European Regional Development Fund (ERDF) via
Welsh Government. The DiRAC Data Intensive service at Leicester is operated 
by the University of Leicester IT Services, which forms part of the STFC DiRAC HPC Facility (www.dirac.ac.uk). The DiRAC Data Intensive service equipment at Leicester was funded by BEIS capital funding via STFC capital
grants ST/K000373/1 and ST/R002363/1 and STFC DiRAC Operations grant ST/R001014/1. 
DiRAC is part of the National e-Infrastructure.

{\bf Open Access Statement-} For the purpose of open access, the authors have applied a Creative Commons Attribution (CC BY) licence to any Author Accepted Manuscript version arising. 
\end{acknowledgement}

%%%%%%%%%%%%%%%%%%%%%%%%%%%%%%%%%%%%%%%%%%%%%%%%%%%%%%%%%%%%%%%%%%%%%%%%%%%%%
\bibliography{references}

\end{document}